\documentclass[12pt,draftclsnofoot,onecolumn]{IEEEtran}  
\usepackage{graphicx,graphics,amssymb,amsmath,here}

\newtheorem{rem}{\bf Remark}

\title{\LARGE \bf
A New Identification Framework For Off-line Computation of Moving-Horizon Observers}

\newtheorem{prop}{Proposition}

\author{Mazen  Alamir
\thanks{Mazen Alamir is with CNRS, Gipsa-lab, University of Grenoble. 11, Rue des Math\'{e}matiques, 38402, Saint-Martin d'H\`{e}res, France. 
        {\tt\small mazen.alamir@grenoble-inp.fr}}%
}

\begin{document}

\maketitle
\thispagestyle{empty}
\pagestyle{empty}

\begin{abstract}
\noindent In this paper, a new nonlinear identification framework is proposed to address the issue of off-line computation of moving-horizon observer estimate. The proposed structure merges the advantages of nonlinear approximators with the efficient computation of constrained quadratic programming problems. A bound on  the estimation error is proposed and the efficiency of the resulting scheme is illustrated using two state estimation examples. 
\end{abstract}
\section{INTRODUCTION}\label{sec-Intro}
\noindent  State estimation is a key issue in nonlinear systems control and diagnosis. Algorithms that achieve this task are called observers. These algorithms attempt to reconstruct the evolution of the state vector by using the only measured (and generally noisy) quantities.  As far as nonlinear systems are concerned, many observation techniques have been developed during the last four decades. This includes high-gain observers \cite{Gauthier92}, sliding-modes observers \cite{Slotine1987}, Moving-Horizon Observers (MHOs) \cite{Michalska1995} and naturally, the widely used Extended-Kalman-Filter (EKF). Excellent reviews of nonlinear observer design techniques can be found in \cite{Simon06} and \cite{Summer_school07}. \\ \ \\ 
Amongst all possible observer design alternatives, MHO technique has witnessed an increasing interest these last years because of its ability to handle constraints and to fully exploit precise and generally nonlinear models of the dynamic processes under study. This observer requires on-line solution of non convex optimization problems in which the cost function is the integral output prediction error while the decision variable is the set of unknown quantities to be recovered (state and unknown parameter vectors). \ \\
Despite encouraging recent advances on the issue of real-time computation of MHOs (see \cite{Alessandri2010,Diehl09} and the references therein for a recent survey on the topics), the highly demanding on-line computation involved may question the feasibility of the algorithm when system needing high sampling rate are involved or when the use of highly involved optimization software is prohibited by real-life context. When the combination of such obstacles and the absence of mathematical structure that renders impossible the use of analytical observers, something has to be done in order to achieve the estimation task. \\ \ \\ 
The ideas proposed in this paper follow a suggestion made by \cite{Alessandri08} aiming at identifying off-line the relationship between the sequence of (input/output) measurements and the corresponding initial state at the beginning of the moving observation window.  By doing so, the cumbersome on-line optimization step involved in Moving-Horizon Observers \cite{Michalska1995}  (MHOs) can  be avoided. Concrete implementation of such an idea using Neural Networks (NNs) identification structure has been proposed in \cite{Alessandri11}. As pointed out in \cite{Alessandri08},  NNs structures, like all standard nonlinear approximators offer high approximation capabilities at the price of non convex optimization schemes which generally suffer from convergence and computation time issues. In the present paper, a novel nonlinear identification scheme is proposed, which after a suitable change in the decision variables, can be solved using constrained quadratic programming.  Such a feature is crucial  if this optimization problem lies in the inner loop of a whole static game optimization formulation (such as the one proposed in \cite{Alessandri08}) that may be needed to assess the convergence of the resulting MHO. More precisely, the solution of the static game is needed to compute the set of approximator parameters that achieves sufficiently small maximum error over all possible initial states (and initial estimation error). The price to pay in order to gain computational efficiency is a theoretical restriction of the class of situations that can be addressed as the proposed identification structure possesses no universal approximation property. For this reason, the comparison between the two identification schemes is problem-dependent. \\  \ \\
It is important to underline that in the scheme proposed in \cite{Alessandri08,Alessandri11} a set of non convex optimization problems are solved off-line (using different initial conditions and initial state estimation errors) in order to built the learning data for the identification step. Instead,  the scheme of the present paper avoids the use of such non convex optimization by focusing on the model-based state/output relationship that can be discovered using system simulation providing the learning data followed by an identification step using specific class of nonlinear structures. This so identified function gives the output-related guess of the state which can then be amended by a model related term in a rather trivial way in order to recover a kind of classical measurement/model confidence trade-off.  By doing so, one can avoid the risk of local minima that may corrupt the quality of the learning data. \\ \ \\  
The paper is organized as follows. First, section \ref{secprobform} defines the state estimation-related identification problem. Section \ref{secrecall}  describes the proposed nonlinear identification setting and gives a bound on the state estimation error. The way the learning data used in the identification scheme is built is shown in section \ref{seclearning}. Illustrative examples are given in section \ref{secexamples} while section \ref{secconc} summarizes the contribution of the paper and suggests hints for further investigations.  
\section{State estimation as an identification problem}\label{secprobform}
\noindent Let us consider a nonlinear system given by:
\begin{eqnarray}
x(k+1)=f(x(k),u(k)) \quad;\quad y(k)=h(x(k),u(k)) \label{syst1}
\end{eqnarray}  
where $x\in \mathbb{R}^{n}$, $u\in \mathbb{R}^{n_u}$ and $y\in \mathbb{R}^{n_y}$ represent the state, the measured input and the measured output vectors respectively. The integer $k$ refers to the sampling instant $k\tau$ for some sampling period $\tau>0$.  Regardless of the algorithm that may be used to reconstruct the state vector using the measured quantities, the implicit assumption that underlines the possibility of state reconstruction is that there is an integer $N\in \mathbb{N}$ and a map $\mathcal{F}$ such that the following approximation holds: 
\begin{eqnarray}
x(k)\approx \mathcal{F}(Z(k)) \label{basicfunction} 
\end{eqnarray} 
where $Z(k)$ is the regressor built up with the past measured quantities according to:
\begin{eqnarray}
Z(k):= \begin{pmatrix}
\tilde{y}(k)\cr \tilde{u}(k)
\end{pmatrix}\in \mathbb{R}^{N(n_u+n_y)=:n_z} \label{defdeZ} 
\end{eqnarray} 
where:
$\tilde{y}(k):=(y^T(k),\cdots,y^T(k-N+1))^T$ and $\tilde{u}(k):=
u^T(k),\cdots,u^T(k-N+1))^T$. 
Note that in the absence of measurement noise, the implicit definition of the map $\mathcal{F}$ involved in (\ref{basicfunction}) is given by the solution of the following optimization problem:
\begin{eqnarray}
&&\mathcal{F}(Z(k)):=X(k,k-N+1,x^*,\tilde{u}(k))\label{ygtf1}\\ 
&&x^*:=\mbox{\rm arg}\min_{x} J(x,Z(k)):=\sum_{i=0}^{N-1}\left\|y(k-i)-Y(k-i,k-N+1,x,\tilde{u}(k))\right\|^2 \label{ygtf2} 
\end{eqnarray}  
where $X(j,k-N+1,x,\tilde{u}(k))$ [resp. $Y(j,k-N+1,x,\tilde{u}(k))$] refers to the model-based predicted value at instant $j$ of the state [resp. output] when the state at instant $k-N+1$ is equal to $x$ and when the sequence of controls defined by $\tilde{u}(k)$ is applied on the time interval $[k-N+1,k]$. \\
It results that if one obtains through off-line computations a good approximation of the map $\mathcal{F}$ involved in (\ref{basicfunction}), then the measurement-related part of the MHOs would be obtained without on-line optimization. This is obviously a static identification problem. More precisely, since this identification is to be obtained for each component of the state vector one needs to reconstruct, the basic generic problem one has to solve is the one consisting of finding a nonlinear map $F$ that links a scalar quantity $r$ (a component of the state vector) to a regression vector $Z$, namely: 
\begin{eqnarray}
r\approx F(Z)\quad;\quad r\in \mathbb{R}\quad;\quad Z\in \mathbb{R}^{n_z} \label{jnhbgv} 
\end{eqnarray} 
\begin{rem}
\noindent Note that, as suggested in \cite{Alessandri08}  the cost function involved in (\ref{ygtf2}) may contain a regularization term $\|x-\bar{x}\|^2$ where $\bar{x}$ stands for the predicted value of the decision variable based on the state space model and the last estimate. This enables a regularization of the estimation process and enhance more stability of the iterates. Such regularization can be decoupled from the identification process by introducing afterward the following final estimation which represents a trade-off between the measurement related estimation and the model related estimation based on the previous solution:
\begin{eqnarray}
\hat{x}=\lambda \mathcal{F}(Z)+(1-\lambda)\bar{x} \label{defdelambda} 
\end{eqnarray} 
where $\mathcal{F}(Z)$ is the measurement-based identified part while $\bar{x}$ is the model predicted part based on the past value of the estimation. In the sequel, we focus on $\mathcal{F}(Z)$ term which corresponds to the choice $\lambda=1$ in (\ref{defdelambda}) . 
\end{rem}

\section{A Nonlinear Identification Framework}\label{secrecall} 
\noindent Identification of nonlinear relationships is an open issue. Many frameworks have been proposed including neural networks, Wiener-Hammerstein, Volterra Series based formulations \cite{DoyleIII2001} to cite but few possibilities. While offering high approximation capabilities, nonlinear approximators need non convex optimization in order to compute the approximator parameters. Such optimization problems suffer from computation time issue and the presence of local minima that may prevent the solver from reaching the global minimizers. In this paper, a nonlinear approximator is proposed that can be computed through constrained QP formulation at the price of lesser universality. \ \\ 
More precisely, in this paper, attention is focused on nonlinear maps $F$ that takes the following form: 
\begin{eqnarray}
F(Z):=\Gamma^{-1}(L^TZ)\quad;\quad \Gamma(\cdot)\quad \mbox{\rm strictly increasing}\label{defstructF}  
\end{eqnarray} 
The existence of $\Gamma^{-1}$ is guaranteed by the strict monotonicity of $\Gamma$. Note that putting together (\ref{jnhbgv}) and (\ref{defstructF}) enables the identification problem to be re-formulated as the one of finding: 
\begin{itemize}
\item the vector $L\in \mathbb{R}^{n_z}$ and 
\item the monotonic increasing function $\Gamma$ such that the following approximation holds: 
\begin{eqnarray}
\Gamma(r(k))\approx L^TZ(k) \label{Gammar} 
\end{eqnarray}  
\end{itemize}   
\begin{rem}
\noindent Note that (\ref{defstructF}) is a nonlinear parameterization as one has to find both $\Gamma^{-1}$ and $L$ which operate nonlinearly. Moreover, while this structure is adapted to the derivation of efficiently solvable constrained QP problems, it is not universal in the sense that any function $F(Z)$ cannot necessarily be represented using the structure (\ref{defstructF}).  Only the class of functions $F$ for which there exists a linear combination of the components of $Z$ that maps to $Z$ through a monotonic function is eligible. This property is impossible to check a priori, but the efficiency of the QP problem solvers makes it easy to check even for very rich parameterization. If the residual is still too high, then the system is surely out of this class since failure cannot result from local minima as it is the case in standard nonlinear approximators computation. 
\end{rem}

The general form of the l.h.s of (\ref{Gammar}) that is used hereafter is the one given by: 
\begin{eqnarray}
\Gamma(r)&:=&\Bigl[B\bigl(\dfrac{r-r_{min}}{r_{max}-r_{min}}\bigr)\Bigr]\cdot \mu=:\Bigl[B(\eta(r))\Bigr]\cdot \mu=\sum_{j=1}^{n_b}\mu_j\Bigl[B^{(j)}(\eta(r))\Bigr] \label{plokrd} 
\end{eqnarray}  
where $r_{min}$ and $r_{max}$ are the minimum and maximum values of $r$ over the learning data (see section \ref{seclearning}) while $B(\eta)$ is a function basis that is hereafter defined according to: 
\begin{eqnarray}
\Bigl\{B^{(j)}\Bigr\}_{j=1}^{n_b}:=\Bigl\{1\Bigr\}\cap \Bigl\{B_1^{(i)}\Bigr\}_{i=1}^{n_m-1}\cap \Bigl\{B_2^{(i)}\Bigr\}_{i=1}^{n_m} \label{efdelabase} 
\end{eqnarray} 
where the number of functions in the set is $n_b=2n_m$ while the functions $B_1^{(i)}$ and $B_2^{(i)}$ are defined by: 
\begin{eqnarray*}
B_1^{(i)}:=(1+\alpha_i)\dfrac{\eta}{1+\alpha_i\eta} \quad;\quad 
B_2^{(i)}:=\dfrac{\eta}{1+\alpha_i-\alpha_i\eta} \label{defdesB2}
\end{eqnarray*} 
The coefficients $\alpha_i$ are given by $\alpha_i:=\exp\bigl(\beta(1-i)\bigr)-1$.
Note that many other function basis can be used although the author's experience suggests that the function basis proposed above is rather appropriate given the monotonicity character of the targeted nonlinear map. \\
Now by combining (\ref{Gammar})-(\ref{plokrd}) , it follows that the unknowns $L$ and $\mu$ may be obtained by solving the following least squares problem:
\begin{eqnarray}
\min_{L,\mu}\sum_{(r,Z)\in \mathcal{E}} \|B(\eta(r))\mu-Z^TL\|^2 \quad \mbox{\rm under the constraint (\ref{matrixineq})} \label{defdeLSopt} 
\end{eqnarray} 
where $\mathcal{E}$ is the learning data including a large number of instantiations of the pairs $(r,Z)$. Note however that the least squares minimization invoked in (\ref{defdeLSopt}) has to be done taking into account the following constraints: \\
1) $\mathbf{\Gamma}$ {\bf  must be strictly increasing}. This leads to the following inequality constraint on the parameter vector $\mu$:
\begin{eqnarray}
\forall \eta\in [0,1]\quad\Bigl[\dfrac{dB}{d\eta}(\eta)\Bigr]\mu\ge \varepsilon \label{contr1} 
\end{eqnarray} 
for some a priori chosen lower bound of the derivative $\varepsilon>0$.\\
2) The following {\bf normalization integral constraint has to be satisfied} in order to avoid trivial zero values trivial solution ($L=0$, $\mu=0$):
\begin{eqnarray}
\Bigl[\int_{0}^1 B(\eta)d\eta \Bigr]\cdot \mu=\dfrac{1}{2}\Bigl[r_{min}+r_{max}\Bigr] \label{contr2} 
\end{eqnarray} 
\begin{rem}
Note that the use of $(r_{min}+r_{max})/2$ in the r.h.s of (\ref{contr2}) is arbitrary since the solution of (\ref{defdeLSopt}) is defined up to a multiplicative gain.  Note however that the r.h.s of (\ref{contr2}) is inspired by the particular case where a linear function fits the learning data. In this case, $\Gamma$ can be taken to be the identity map and in this case, the integral has to equal the mean value of $r$. 
\end{rem}
 \ \\  
The two sets of constraints (\ref{contr1}) and (\ref{contr2}) are affine in the decision variable $\mu$. Therefore  by considering a sequence of values $0=\eta_1<\eta_2<\dots<\eta_q=1$, these constraints can be approximated by the following matrix inequalities: 
\begin{eqnarray}
[A_{ineq}]\mu\le B_{ineq}\in \mathbb{R}^q\quad;\quad [A_{eq}]\mu=B_{eq}\in \mathbb{R} \label{matrixineq} 
\end{eqnarray} 
To summarize, the quadratic function (\ref{defdeLSopt}) in the decision variables $L$ and $\mu$ is minimized under the linear constraints (\ref{matrixineq}) to obtain the optimal parameters $L_{opt}$ and $\mu_{opt}$. \ \\
Note that there is no loss in generality in taking $\Gamma$ strictly increasing since it suffices to take $L$ of opposite sign to make (\ref{Gammar}) valid for a strictly decreasing map.  \\ \ \\  
The following result on the estimation error is straightforward: 
\begin{prop} {\bf If}  Let $\mathbb X\subset \mathbb{R}^{n}$ be a subset of the state space to which belong all state of interest. The following conditions are satisfied: \\
1) The admissible control sequences are bounded (i.e. $\tilde{u}\in \tilde{\mathbb{U}}$) and lead for any $x^{(0)}\in \mathbb{X}$ to a bounded sequence of output ($\tilde{y}\in \tilde{\mathbb{Y}}$)\\
2) there is an upper bound $\varepsilon_x>0$ on the identification residual:
\begin{eqnarray}
\sup_{(x^0,\tilde{u})\in \mathbb{X}\times \tilde{\mathbb U}}\left\| x^{(0)}-\mathcal F\bigl(Z(x^{(0)},\tilde{u})\bigr)\right\|\le \varepsilon_x \label{defdevarepsilon} 
\end{eqnarray} 
3) For any initial state $x^{(0)}$, the combined effect of noise and model mismatches on the regressor $Z$ is bounded according to:
\begin{eqnarray}
\|Z_{\mbox{\rm reel}}(x^{(0)},\tilde{u}) -Z(x^{(0)},\tilde{u})\|_\infty\le \gamma \label{defdegamma} 
\end{eqnarray} 
where $Z_{\rm reel}$ denotes the real measurement matrix (which is not obtained by simulation) that may be obtained on the real system starting from $x^{(0)}$ and under the control sequence $\tilde u$.\ \\ \ \\ 
{\bf then} the estimation error is of the form: 
\begin{eqnarray}
\| \hat{x}-x\| \le \varepsilon_x+O(\gamma)
\end{eqnarray}  
\end{prop}
{\sc Proof}. First of all, note that assumption 2) implies that all nominal measurement regressors $Z$ of interest belong to some compact set $\mathbb{Z}$. Using the assumptions (\ref{defdevarepsilon})-(\ref{defdegamma}), one clearly has: 
\begin{eqnarray*}
\left\| x^{(0)}-\hat{x}^{(0)}\right\|\ &&:=\left\| x^{(0)}-\mathcal F\bigl(Z_{\mbox{reel}}(x^{(0)},\tilde{u})\bigr)\right\|\nonumber \\ 
&&\le \|x^{(0)}-\mathcal F(Z)\|+\|\mathcal F(Z)-\mathcal F(z_{\rm reel})\|\nonumber \\
&&\le \varepsilon_x+\max_{Z\in \mathcal{Z}}\left\|\dfrac{\partial \mathcal{F}}{\partial Z}(Z)\right\|\cdot \|Z-Z_{\mbox{\rm reel}}\|\le \varepsilon_x+\sqrt{n_z}\cdot M(\gamma)\cdot \gamma
\end{eqnarray*}
where $\mathcal{Z}:=\mathbb{Z}+B(0,\gamma)$ and where $M(\gamma)$ is the maximum value of the continuous (by construction) map $\dfrac{\partial \mathcal F}{\partial Z}$ over the compact set $\mathcal Z$. $\hfill \Box$ 
\begin{rem}
Note that the upper bound $\varepsilon_x$ involved in (\ref{defdevarepsilon}) can be used as a cost function $\varepsilon_x(N,n_m,\beta)$ to be minimized in the decision variable $(N,n_m,\beta)$ which are the parameters of the approximator. This obviously results in a static game in which the identification step appears in the inner loop. This strengthens the relevance of having easy to solve identification problem through the constrained QP formulation. Note also that the value of $\varepsilon_x$ reflects to which extent the map linking the regressor $Z$ and the initial state is far from the set of maps described by the structure (\ref{defstructF}). This is because no identification error can be affected to the optimization process as the underlying problem is a quadratic programming one. 
\end{rem}

\section{BUILDING THE LEARNING DATA}\label{seclearning} 
\noindent Assume without loss of generality that one is focused on the identification problem associated to the estimation of $r=x_i$ for some $i\in \{1,\dots,n\}$. Let us also assume that the set of relevant values of the state vector $x$ is contained in some hypercube, namely: 
\begin{eqnarray}
x\in \mathbb{X}:= \displaystyle{\Pi_{i=1}^{n}}[x_i^{min},x_i^{max}]
\end{eqnarray} 
The learning data set $\mathcal{E}$ involved in the definition of the least squares problem (\ref{defdeLSopt}) is obtained through the following steps: \\
1) First, a set of initial states $\bigl\{x^{(j)}_0\bigr\}_{j=1}^{n_g}$
is chosen. This can be obtained using uniform grid on each interval $[x_{i}^{min},x_{i}^{max}]$ of possible values of the $i$-th component of the state.\\
2) A set of $n_g$ control profiles $\bigl\{\tilde{u}^{(j)}\bigr\}_{j=1}^{n_g}\quad;\quad \tilde{u}^{(j)}\in \mathbb{R}^{N_s\cdot n_u}$
is also generated. Each profile defines a control sequence over $N_s>N$ sampling periods. \\
3) For each pair $(x_{0}^{(j)},\tilde{u}^{(j)})$, the system is simulated over $N_{s}$ sampling periods in order to generate the corresponding  state profiles:
\begin{eqnarray}
\Bigl\{X(k,0,x_0^{(j)},\tilde{u}^{(j)})\quad,\quad k\in \{1,\dots,N_s\}\Bigr\}_{j=1}^{n_g}
\end{eqnarray} 
4)  The data described above enables, for each pair $(j,k)\in \{1,\dots,n_g\}\times \{N,\dots,N_s\}$ to obtain: 
\begin{itemize}
\item two sequences $\tilde{y}^{(j)}(k)$, $\tilde{u}^{(j)}(k)$ that defines $Z^{(j)}(k)$ according to (\ref{defdeZ}).
\item the corresponding value of $r^{(j)}(k)$ according to 
$$r^{(j)}(k):=X\Bigl(N,0,X(k-N,x_0^{(j)},\tilde u^{(j)}),\tilde{u}^{(j)}\Bigr) $$
\end{itemize} 
Finally, the learning data set $\mathcal{E}$ involved in the definition of the least squares problem (\ref{defdeLSopt}) is given by: 
\begin{eqnarray}
\mathcal{E}:=\Bigl\{\bigl(r^{(j)}(k),Z^{(j)}(k)\bigr)\Bigr\}_{(j,k)\in \{1,\dots,n_g\}\times \{N,\dots,N_s\}}
\end{eqnarray} 
which is obviously a discrete set of cardinality $n_E$ that is given by: 
\begin{eqnarray}
n_E:=n_g\cdot (N_s-N+1) \label{defdenE} 
\end{eqnarray} 
\section{ILLUSTRATIVE EXAMPLES}\label{secexamples} 
\noindent In this section, illustrative examples are proposed in order to give concrete instantiations of the different steps of the proposed framework.
\subsection{Example 1}
\begin{figure}
\begin{center}
\includegraphics[width=0.6\textwidth]{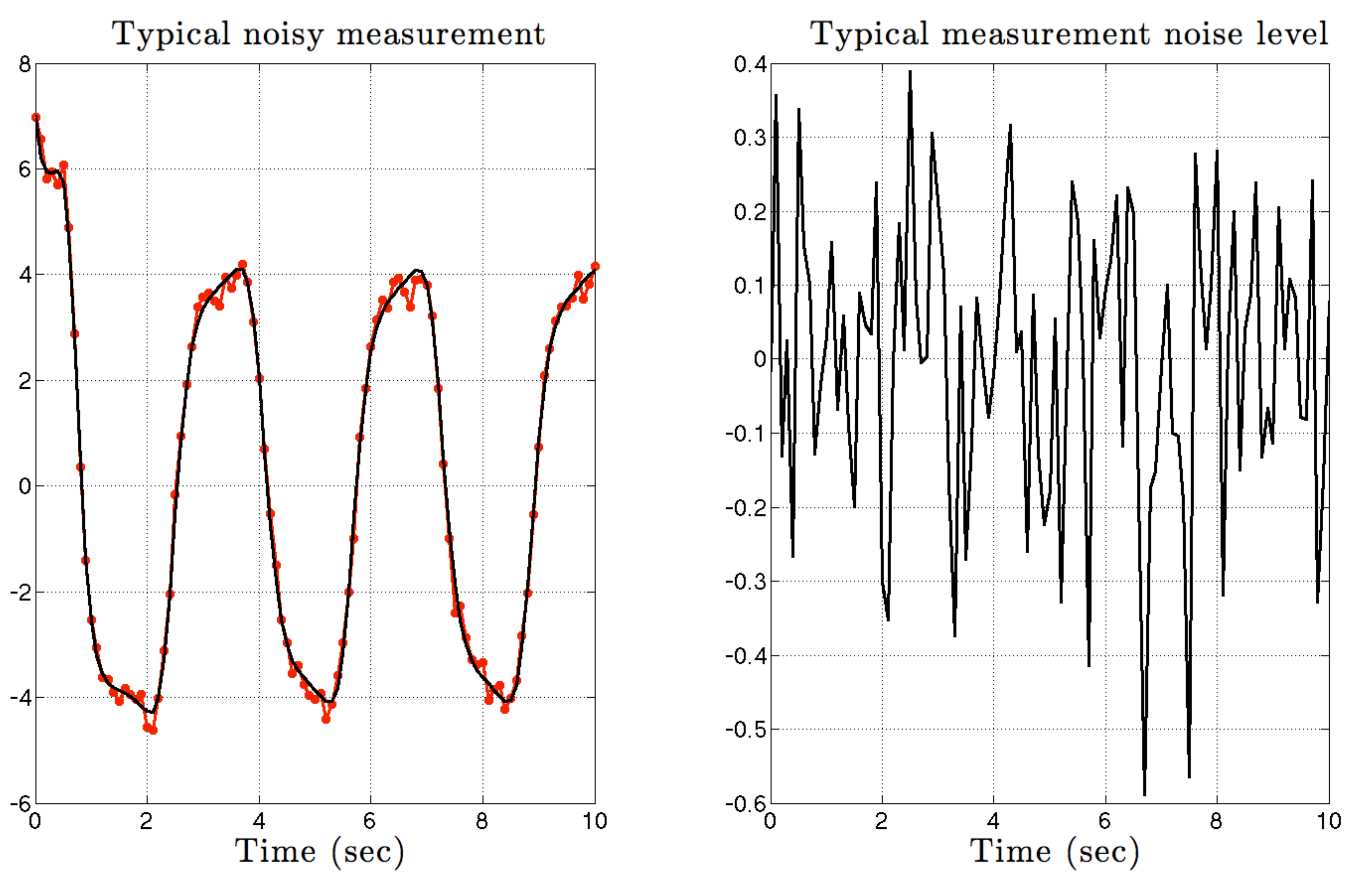}
\end{center}
\caption{Example 1. Typical behavior of the measured output and the corresponding noise. The Gaussian noise used in (\ref{uhygtfrd}) is generated with a variance $\sigma=0.2$}\label{Typical_noise} 
\end{figure}

\begin{figure}
\begin{center}
\includegraphics[width=0.49\textwidth]{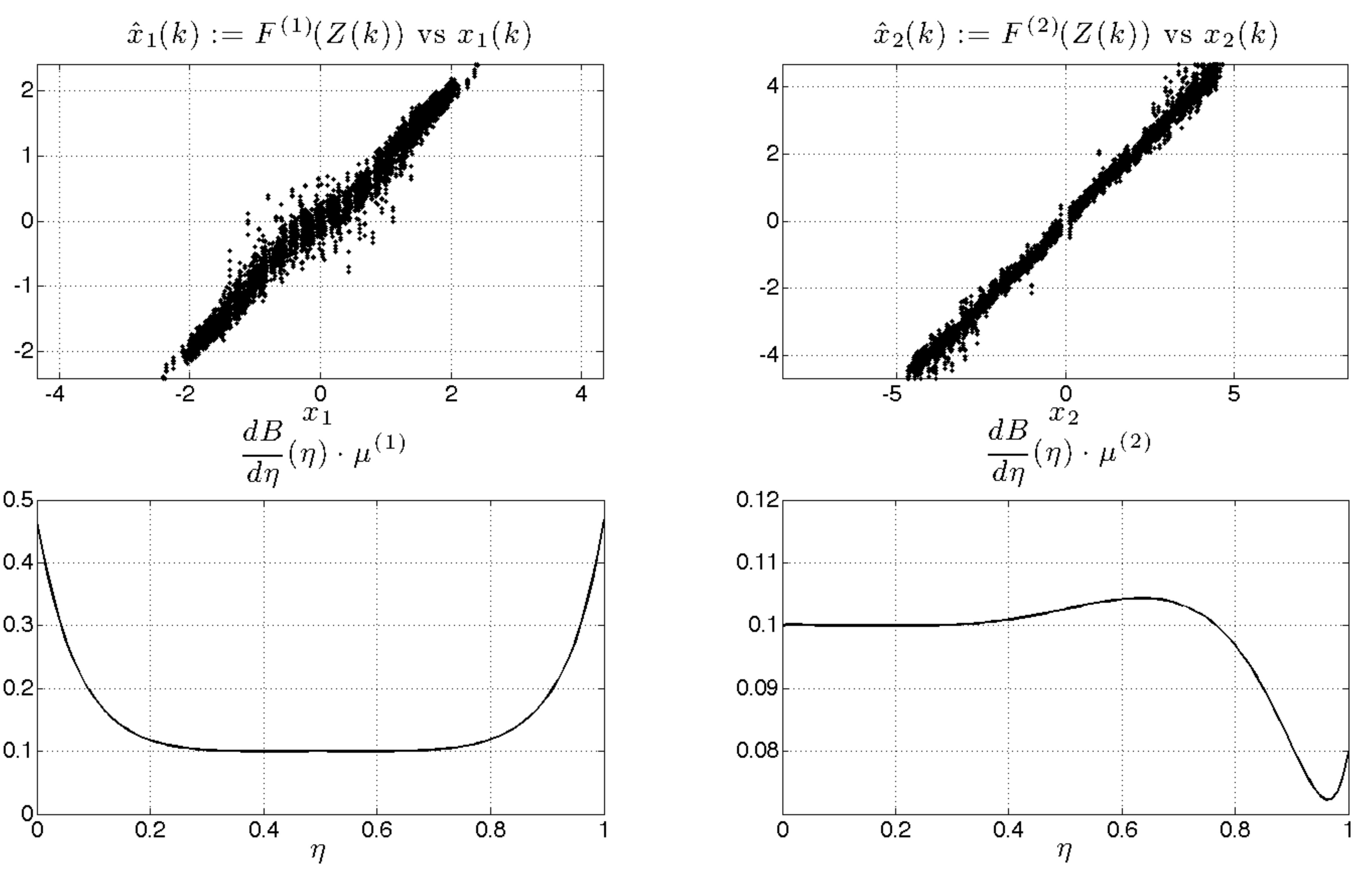}
\includegraphics[width=0.49\textwidth]{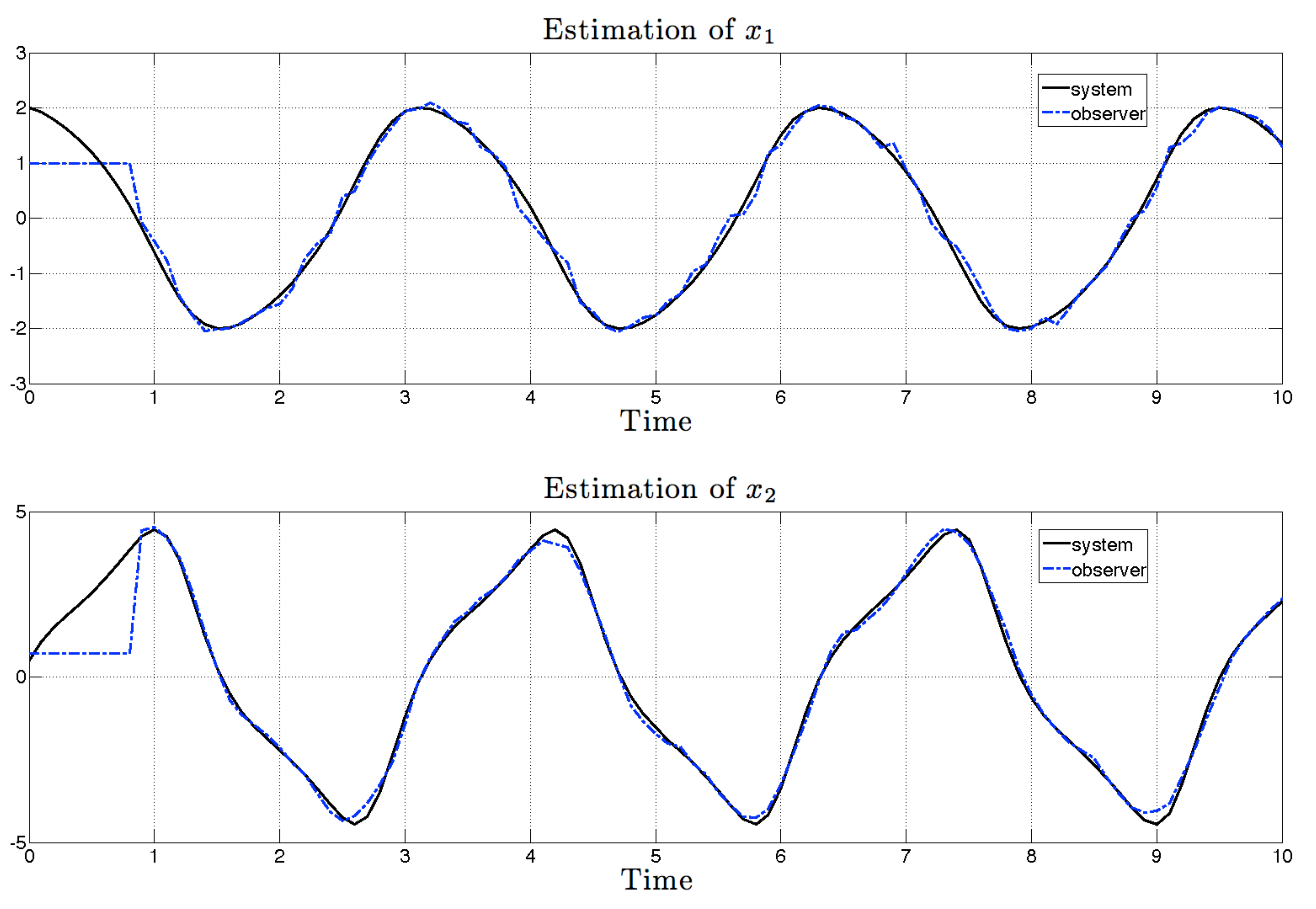}
\end{center}
\caption{Example 1. (Left): A picture showing the results of the nonlinear  identification procedure using a learning set of cardinality $n_E=1456$. The upper plots compares the estimated states based on the identified maps $F^{(1)}$ and $F^{(2)}$ while the lower plots show the gradient of the corresponding monotonic functions $\Gamma^{(1)}$ and $\Gamma^{(2)}$ (Right) : Typical results of the state estimation using a purely output-related estimation. The observer begins once the first $N=10$ first measurements are acquired (first correction at $t=0.9\ sec$). This estimation is obtained while a Gaussian measurement noise $w$ of variance $\sigma=0.2$ is injected (see figure \ref{Typical_noise} for a typical behavior of the noise)}\label{diagonals} 
\end{figure}

\noindent Let us consider the famous Van der Pol oscillator which is governed by:
\begin{eqnarray}
\dot{x}_1=-x_2\quad;\quad \dot{x}_2=4x_1+(1-x_1^2)x_2\quad;\quad y:=x_1+x_2+w \label{uhygtfrd} 
\end{eqnarray} 
where $w$ is a measurement noise assumed here to be white, Gaussian with variance $\sigma=0.2$.  Typical behavior of the resulting noisy measurement is shown in Figure \ref{Typical_noise} (left subplot) together with typical corresponding level of the noise (right subplot). The basic sampling period is taken equal to $\tau=0.1\ sec$. The bounds on the state components leading to the definition of the subset $\mathbb{X}$ are given by  $\mathbb{X}:=[-2,+2]\times [-5,+5] $
which obviously {\em enhances} the nonlinear character of the resulting identification problem (as $2^2$ is not negligible when compared to 1 in the expression of $\dot x_2$). The learning data has been defined using a uniform grid containing $n_g=4^2=16$ initial states. For each state, the system is simulated during $N_s=100$ sampling period ($10\ sec$)  which, according to (\ref{defdenE}) leads to a learning data $\mathcal{E}$ of cardinality $n_E=16\times (100-10+1)=1456$.
Note that since there are two states $x_1$ and $x_2$ to be estimated, two identification problems are solved and two nonlinear maps $F^{(1)}$ and $F^{(2)}$ are obtained for the reconstruction of these two states according to $\hat{x}_i(k)=F^{(i)}(Z(k))$, $i\in \{1,2\}$
The identification parameters $N=10$ and $n_m=5$ are used
which leads to an observation horizon of $N\times \tau=1\ sec$ and a functional basis containing $n_b=10$ elements. \ \\ 
The quality of the resulting matching between the identified and the {\em true} values are shown in the upper subplots of Figure \ref{diagonals} (left plots). Note that the identified values are obtained using an intentionally noised simulation data which enables one to inject the noise already in the identification process resolving the trade-off at this early stage. The lower subplots of Figure \ref{diagonals} (Left) shows the gradient of the resulting nonlinear functions $\Gamma^{(1)}$ and $\Gamma^{(2)}$ that are involved in (\ref{defstructF}) for the two components of the state respectively. One may note the highly nonlinear character of the map $\Gamma^{(1)}$ in particular.  An estimation scenario is shown in Figure \ref{diagonals} (right plots) where the initial state of the system and the observer are respectively given by $x(0)=(2,0.5)^T$ and $\hat{x}(0)=(1,0.7)^T$ and where a new generation of the measurement noise is used (different from the ones used to construct the learning data set). Note that the correction of the observer begins only when data is obtained that covers the observer horizon length. This means that the first correction occurs at $t=(N-1)\tau=0.9\ sec$. It is worth emphasizing here that the observer design is done using only the output measurement related correction in order to concentrate on the contribution of the present paper. It goes without saying that a balanced estimation in which the output-related estimation and the state equation related estimation can be implemented following \eqref{defdelambda} or more generally the standard ideas of \cite{Hasteline2005,Rao2003} and the references therein. 
\subsection{Example 2}
\begin{figure}
\begin{minipage}{0.48\textwidth}
\begin{center}
\includegraphics[width=\textwidth]{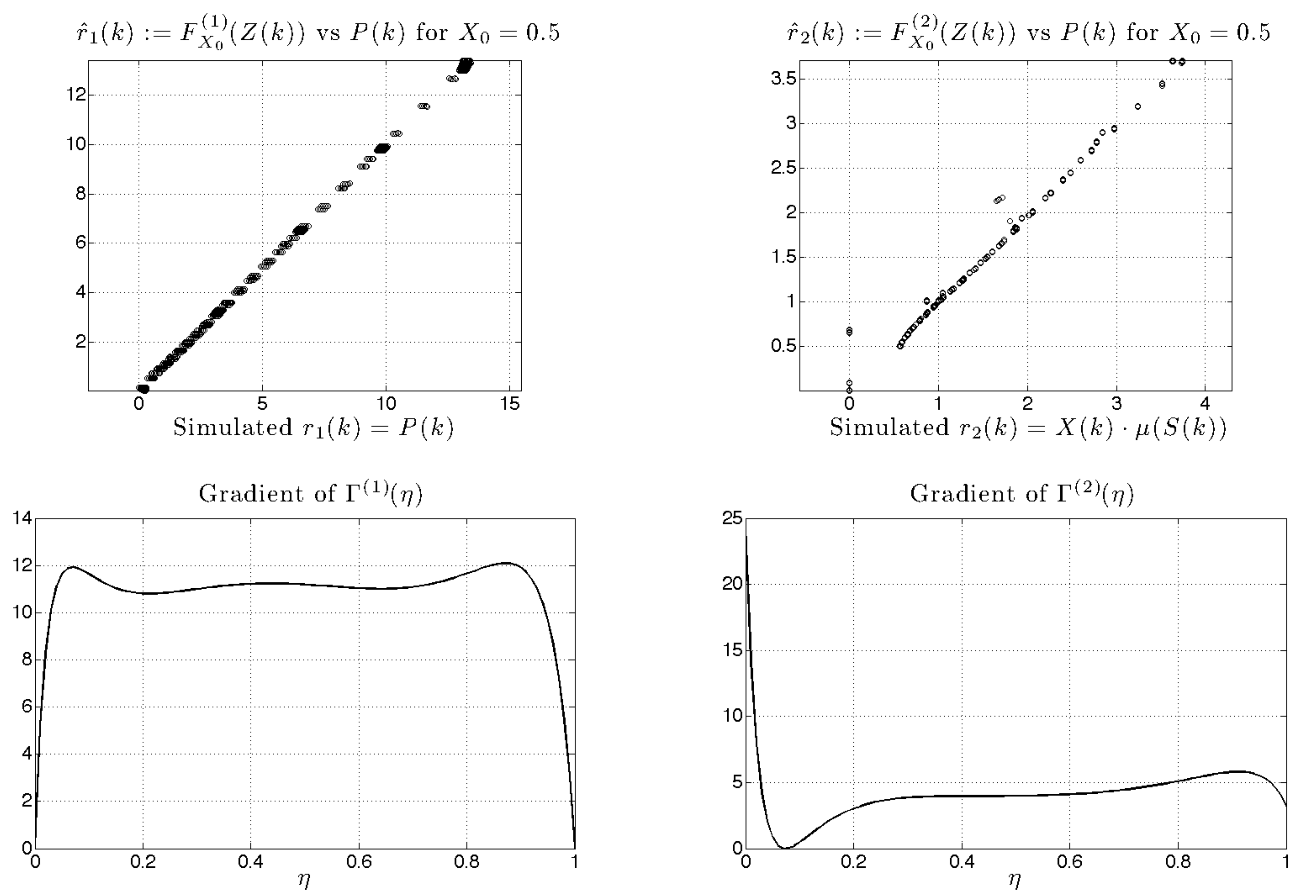}\ \\
(a) Identification results for $X(0)=0.5$
\end{center}
\end{minipage}
\begin{minipage}{0.02\textwidth}
\ 
\end{minipage}
\begin{minipage}{0.48\textwidth}
\begin{center}
\includegraphics[width=\textwidth]{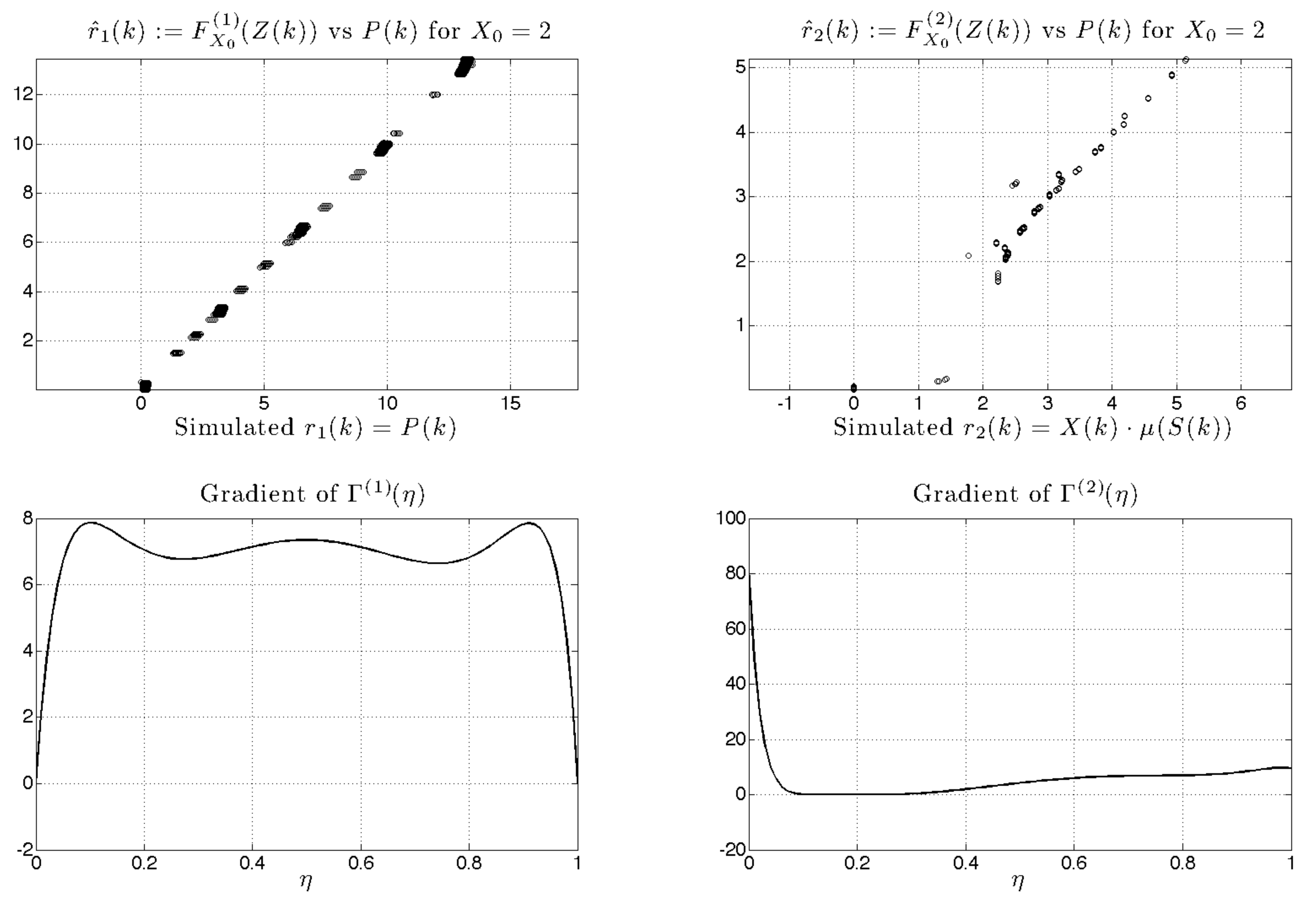}\ \\
(b) Identification results for $X(0)=2.0$
\end{center}
\end{minipage}
\caption{Example 2. Examples of Identification results for $r_1=P$ and $r_2=\mu(S)\cdot X$ for two different values of the E. Coli strain $X_0\in \{0.5, 2\}$.}\label{figEcoliX0} 
\end{figure}

\noindent Let us consider the problem of estimating the state of a dynamic model describing the Escherichia Coli Strain. Many knowledge-based model derivation attempts have been investigated in order to better understand the mechanisms that underline the evolution of the population \cite{Cha2000,Lee1992} or to develop model-based state observers \cite{Nardi2006}.  The dynamic model that is commonly used in deriving dynamic state estimation involves the E. Coli strain $X$ that grows on the limiting substrate $S$ while yielding a final intracellular product: the $\beta$-galactosidae $P$. The model is given by:
\begin{eqnarray}
\dot{X}&=&\mu(S)X-k_d\exp(-\dfrac{k_p}{P})X \label{Ecoli1}\\
\dot{S}&=&-y_s\mu(S) X-k_mX \label{Ecoli2}\\
\dot{P}&=&y_p\mu(S)\dfrac{I}{I+k_I}X-k_d\exp(-\dfrac{k_p}{P})P \label{Ecoli3}   
\end{eqnarray}  
where $\mu$ is the growth rate that is modelled using classical Monod-type relation such as $\mu(S):=\dfrac{\mu_mS}{k_s+S}$
where $\mu_m$ is the maximum specific growth rate for the cell growth (in $h^{-1}$). $k_s$ is the half saturation constant; $k_p$ and $k_d$ are constants involved in the Arrhenius-type death kinetic that depends on $P$. $k_m$ is a maintenance rate that describes the energy required for normal upkeep and repair. $y_s$, $y_l$ [used in the measurement equation (\ref{defdeL}) below]  and $y_p$ are identified coefficients. $I$ stands for the arabinose inducer that is assumed to be constant (no degradation). The output measurement vector is given by $y:=(X+w_1,L+w_2)^T$
where $L$ is the light produced by the bioluminescence that is linked to the state variables by the following expressions: 
\begin{eqnarray}
L=y_l\cdot \mu(S)\dfrac{I}{I+k_l}XP \label{defdeL} 
\end{eqnarray} 
while $w_1$ and $w_2$ are measurement noises that are taken here white, Gaussian and of variances $\sigma_1$ and $\sigma_2$ respectively. The values of the model parameters used in the sequel can be found in \cite{Nardi2006}.  \\ 
\noindent For this example, the framework described above can be used to construct a reduced observer. More precisely, it is shown hereafter that there is a satisfactory solution to the underlined identification problem for any learning set that is constructed using an admissible set of initial conditions of the form $\mathbb{X}(X_0):=\{X_0\}\times [0,5]\times [0,0.3]$ 
with the following parameters $\tau=0.2\ ;\ N_s=40\ ;\ N=3\ ;\ n_m=5\ ;\ n_g=25$.  This means that for each initial value $X_0$ of the E. Coli strain, $n_g=25$ simulations of the system with different initial states (sharing all the same value $X_0$ and different values of $P_0$ and $S_0$) is simulated during $8$ time units ($=N_s\tau=40\times 0.2$) hence generating a learning set of cardinality $n_E$ given by $n_E=25\times (40-3)=925$. Note that two identification problems are to be defined and solved using the following definition of the quantities $r_1$ and $r_2$ to be identified: 
\begin{eqnarray}
r_1&:=&P\quad;\quad r_2=\mu(S)\cdot X
\end{eqnarray} 
Note that if $r_1$ and $r_2$ are well estimated then $\mu(S)$ can also be well estimated since $X$ is assumed to be measured (reduced observer). Note also that since only $\mu(S)$ is involved in the system equation, $S$ is necessarily estimated through $\mu(S)$. \\ 
The identification results are shown on Figure \ref{figEcoliX0} for two different values of $X_0\in \{0.5, 2\}$. One can appreciate that a good match between the estimated $\hat{r}_i$ and the simulated $r_i$ for $i=1,2$ is obtained over the $925$ learning set while using a rather economic parametrization  ($N=3$ and $n_m=5$). This clearly shows that the proposed methodology enables us to perform the estimation scheme provided that the initial value of the state component $X$ is measured at the beginning of the batch. 

\begin{figure}
\begin{minipage}{0.49\textwidth}
\begin{center}
\includegraphics[width=0.9\textwidth]{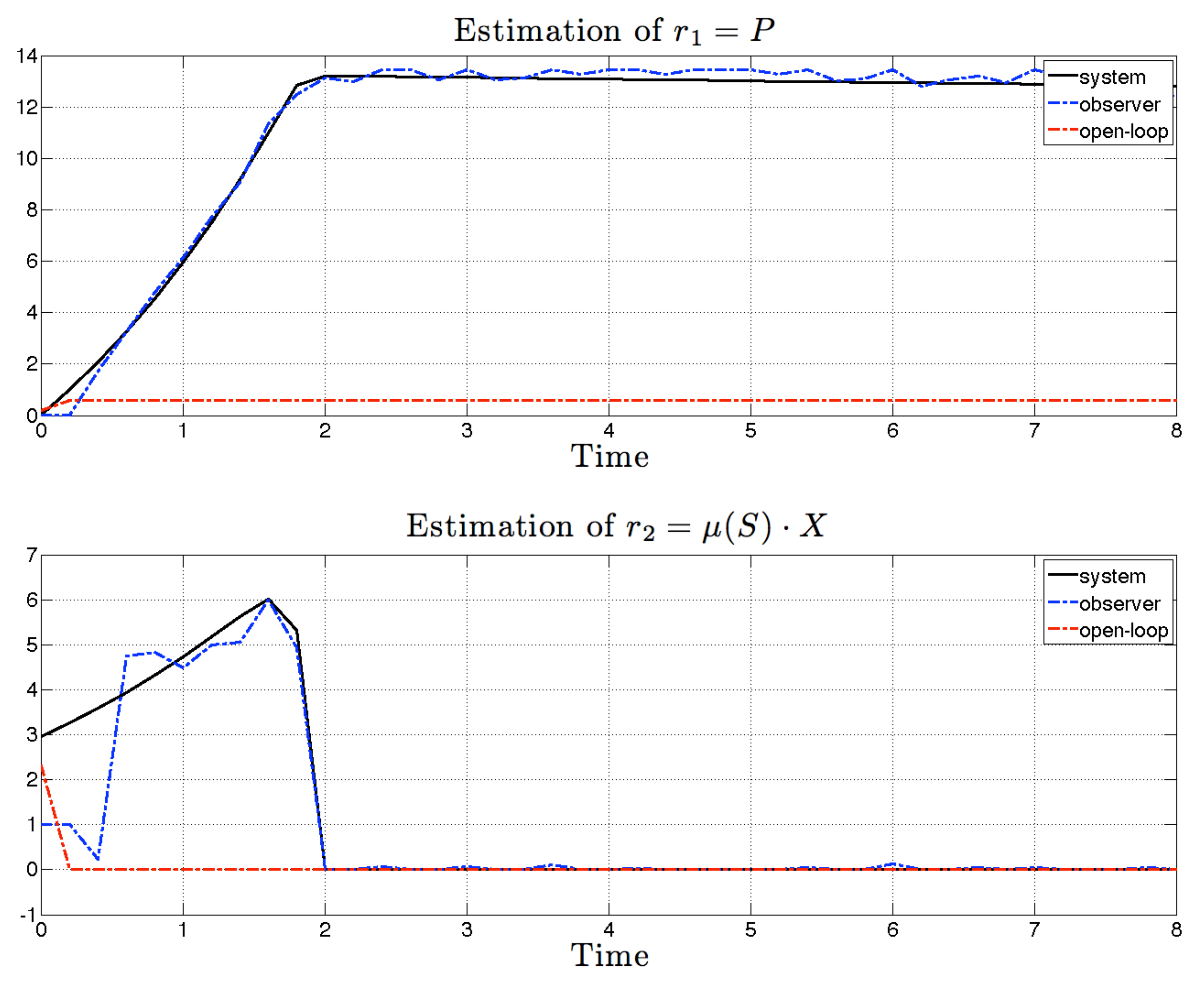}
\end{center}
\caption{Example 2. Behavior of the output-based estimation of $r_1=P$ and $r_2=\mu(S)\cdot X$. The measurement noise variances are respectively given by $\sigma_1=0.1$ and $\sigma_2=20$. Typical behavior of the measurement noise can be observed oin Figure \ref{noiseEcoli}.}\label{estimation5} 
\end{minipage}
\begin{minipage}{0.02\textwidth}
\ 
\end{minipage}
\begin{minipage}{0.49\textwidth}
\begin{center}
\includegraphics[width=0.9\textwidth]{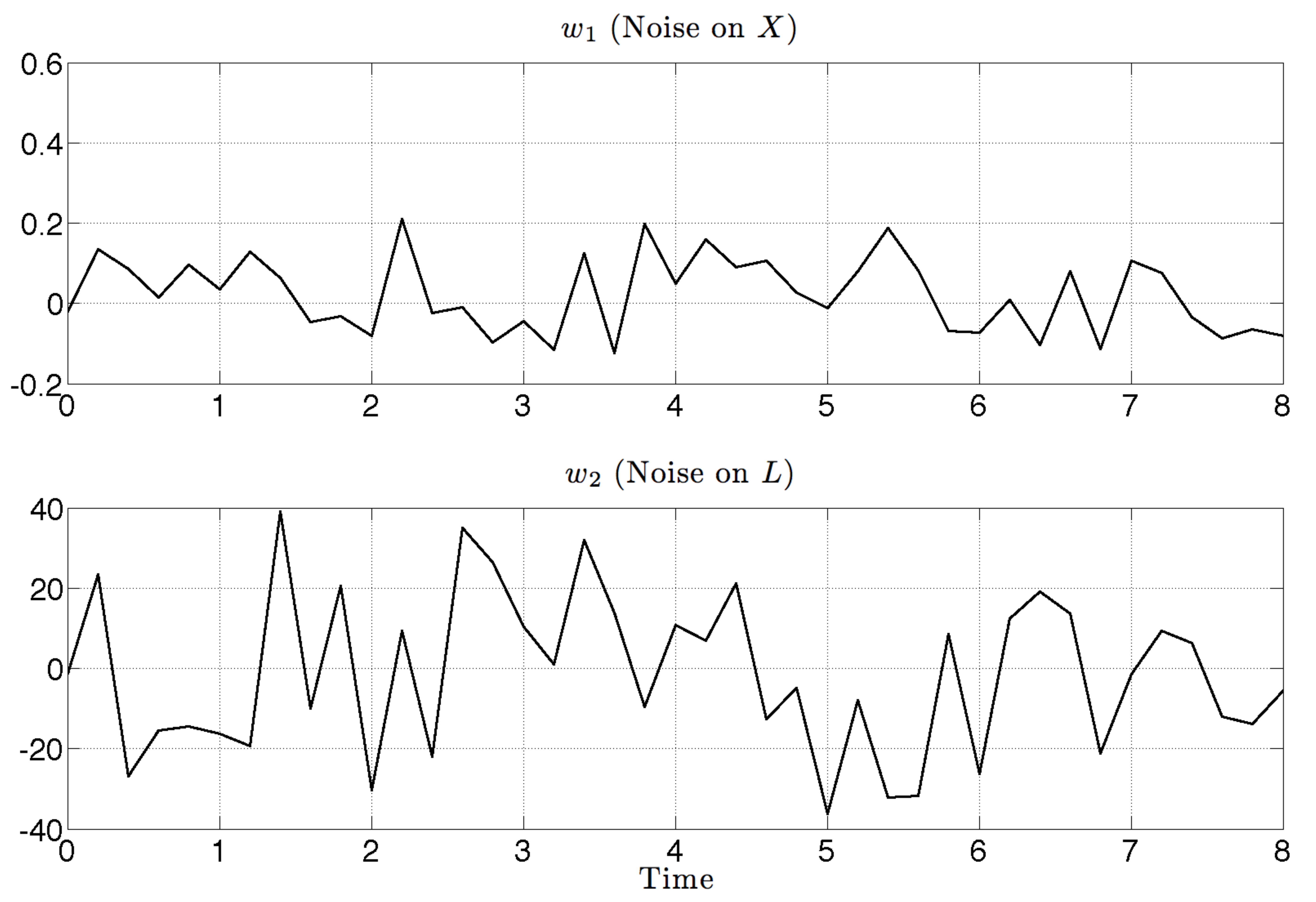}
\end{center}
\caption{Example 2: Typical behavior of the measurement noise used in the simulation of the state estimation depicted in Figure \ref{estimation5}.}\label{noiseEcoli} 
\end{minipage}
\end{figure}

Figure \ref{estimation5} shows typical behavior of the output-based state estimation that used the two nonlinear maps identified above. Note that the  noises $w_1$ and $w_2$ that affect the measured signals used in the construction of the regressor  are given by $\sigma_1=0.1$ and $\sigma_2=20$. This leads to a noise level that can be observed in Figure \ref{noiseEcoli}. 
\section{Conclusion and future work}\label{secconc} 
\noindent In this paper, a nonlinear approximator has been proposed for a class of nonlinear relationships and has been applied in the context of moving-horizon observer design. The proposed scheme offers the advantage of requiring only a constrained QP problem solution and can therefore be efficiently integrated in the inner loop of a global scheme aiming at optimizing the approximator parameters. \ \\ 
A potential research line is to investigate a systematic computation of the optimal triplet $(N,\tau,\beta)$ defining the nonlinear approximator. Indeed, a convenient ({\em optimal}) choice of these parameters is intimately linked to the noise level as well as the uncertainty structure. A good knowledge of the latter is crucial to obtain pertinent choice of these parameters which had been found in this paper by trial and error approach. \ \\ \ \\ Another research track concerns the use of sparse identification techniques in order to derive low dimensional parameter vector. This can be greatly facilitated by the availability of the Lagrange multipliers of the QP problem that underline the identification step.


\end{document}